\documentclass{aa}
\usepackage{aabib99,amssymb,psfig}

\begin{document}

\thesaurus{02     
  (02.09.1;  
   02.13.2;  
   09.10.1;  
   11.10.1)} 

\title{Stability and heating of magnetically driven jets from Keplerian
accretion discs}


\author{F. Thomsen\inst{1}
\and {\AA}. Nordlund\inst{1}}

\institute{The Niels Bohr Institute for
           Astronomy, Physics and Geophysics,
           Juliane Maries Vej 30,
           DK-2100 Copenhagen {\O}, Denmark}

\offprints{$\mathrm{aake@astro.ku.dk}$}

\date{Received date, accepted date}

\maketitle


\begin{abstract}

We have performed
3-D numerical magnetohydrodynamic (MHD) jet
experiments to study the instabilities associated with strongly toroidal magnetic
fields. In contemporary jet theories, a highly wound up magnetic
field is a crucial ingredient for collimation of the flow. If such magnetic
configurations are as unstable as found in the laboratory and by 
analytical estimates, our understanding of MHD jet driving and collimation
has to be revised.

A perfectly conducting Keplerian disc with fixed density, rotational
velocity and pressure is used as a lower boundary for the jet. Initially, the
corona above the disc is at rest, permeated by a uniform magnetic field, and is
in hydrostatic equilibrium in a softened gravitational field from a point mass. The
mass ejection from the disc is subsequently allowed to evolve according to
deviations from the initial pressure equilibrium between disc and corona.

The energy equation is
solved, with the inclusion of self-consistently computed heating by viscous
and magnetic dissipation. We find that magnetic dissipation may have profound
effects on the jet flow as: 1) it turns on in highly wound up magnetic
field regions and helps to prevent critical kink situations; 2) it influences
jet dynamics by re-organizing the magnetic field structure and increasing
thermal pressure in the jet; and 3) it influences mass loading by increasing
temperature and pressure at the base of the jet.

The resulting jets evolve into time-dependent, non-axisymmetric
configurations, but we find only minor disruption of the jets by e.g.\ the kink
instability.

\end{abstract}


\section{Introduction}

Though astrophysical jets have been observed in a rather wide range of
accreting systems, it is generally assumed that the mechanism for
acceleration and collimation is generic \cite[for recent
reviews]{Livio,spruit00}.
Contemporary jet
theories rely on magnetic forces as the jet producing mechanism, either
with the flow emanating from the disc surface and associated with an
open magnetic field structure \cite{BP82} or with the flow emanating from
the disc-star interface and associated with a closed field structure
connecting disc and central star \cite{pringle}. In the former
so-called disc-wind scenario, which is currently receiving most attention,
the driving force, acting to overcome the gravitational pull of the central
object, is typically regarded as a component of the centrifugal force along
the magnetic field. However, adopting an inertial frame of reference the
driving mechanism may equally be interpreted as magnetic \cite{spruit96}.
The inertia of the outflowing gas eventually becomes dynamically important
and the gas stops corotating with the underlying disc. At this point the
magnetic
field, which cannot easily slip through the highly ionized gas, is wound
up in a cork screw manner between the disc and the vertically accelerated and
less quickly rotating outflow. The tension (``hoop stress'') of the
wound-up magnetic field configuration is generally believed to be the
collimating force, but collimation by a poloidal
field has been proposed as well \cite{Spruit} motivated by instability
arguments against the wound-up field scenario.

The assumed large scale magnetic field in the disc-wind scenario may either
be provided by dynamo processes in the disc
\cite{Brandenburg2000,Brandenburg+2000} or
captured from the environment and advected inwards by the accreting matter
\cite{CaoSpruit}. Numerical work has been carried out to investigate situations
in which a significant fraction of the field lines loop back into the
disc \cite{Romanovab,TBR}. These experiments have focused on the
ejection and acceleration mechanisms as well as the evolution of the
magnetosphere close to the disc, and have not followed the flow further
out for extended times.

The numerical work done in relation to the ``classical'' large scale open
field scenario may be divided into
experiments attempting to include (parts of) the accretion disc in the
computational domain \cite{US85,BL95} and more recent work
\cite{Ustyugova,OPS,Meier} which has used the disc only as a fixed lower
boundary
to avoid problems with radial collapse of the disc and thereby relatively
short time evolution of the experiment. Both types of experiments have
been axisymmetric and as such have not questioned the potentially (kink)
unstable wound up magnetic field configuration on which the collimation process
relies. However, Spruit et al.\ \cite*{spruit+00} have recently proposed fast
reconnection processes in more disordered non-axisymmetric large scale magnetic
fields in the jet like gamma ray burst (GRB) scenario. The magnetic
dissipation is proposed for the GRB fireballs e.g.\ to produce the observed
radiation with better efficiency. We find that such magnetic processes may
have severe impact on jet dynamics and stability in particular.

The main purpose of the work presented here is to establish
whether or not the jets in the disc-wind scenario are prone to
catastrophic MHD instabilities.
This calls for an implementation of a setup in three dimensions which will be
described in Sect.~\ref{sec:model}. For simplicity, we assume a large-scale
open magnetic field structure and use the disc as a fixed boundary. The
details of the initial conditions and boundary conditions are found in
Sect.~\ref{sec:ic} and \ref{sec:bc} respectively. Results
are presented in Sect.~\ref{sec:results}, with Sect.~\ref{sec:dynamics}
concentrating on the jet dynamics and Sect.~\ref{sec:stability} on
the observed 3-D jet stability in the experiments.
In Sect.~\ref{sec:discussion} we discuss the results with special emphasis on
thermal properties and magnetic field structure. Finally, the paper
is summarized and conclusions are presented in Sect.~\ref{sec:conclusion}.


\section{Model}
\label{sec:model}

The jet flow is described numerically by solving the MHD
equations in the following form;
\begin{equation}
\frac{\partial \rho}{\partial t} = - \nabla \cdot \rho \mathbf{u},
\end{equation}
\begin{equation}
\frac{\partial(\rho \mathbf{u})}{\partial t} =
- \nabla \cdot \left( \rho \mathbf{uu} - \tau^\prime \right) + \nabla P +
\mathbf{J \times B} - \rho \nabla \Phi,
\end{equation}
\begin{equation}
\mu \mathbf{J} = \nabla \times \mathbf{B},
\end{equation}
\begin{equation}
\mathbf{E} =  \eta \mathbf{J} - \mathbf{u \times B},
\end{equation}
\begin{equation}
\frac{\partial \mathbf{B}}{\partial t} = - \nabla \times \mathbf{E},
\end{equation}
\begin{equation}
\frac{\partial e}{\partial t} =
- \nabla \cdot e \mathbf{u} - P \nabla \cdot \mathbf{u} + Q,
\end{equation}
where $\rho$ is the mass density, $\mathbf{u}$ the velocity vector,
$\tau^\prime$ the viscous stress tensor, $P$ the thermal gas pressure,
$\Phi$ the gravitational potential, $\mathbf{B}$ the magnetic flux density
vector, $\mathbf{J}$ the electric current density vector,
$\mathbf{E}$ the intensity of the electric field, $\eta$ the electric
resistivity, $e$ the internal energy per unit volume and $Q$ the sum of
viscous and Joule dissipation.

\subsection{Dimensions and numerics}

\begin{table*}[b!t]
\begin{center}
\begin{tabular}{l|c|c|c}
\hline
\multicolumn{1}{c|}{\textbf{Quantity}} &
\multicolumn{1}{c|}{\textbf{Unit}} &
\multicolumn{1}{c|}{\textbf{YSO}} &
\multicolumn{1}{c}{\textbf{AGN}} \\
\hline \hline
Velocity                 & $\sqrt{ GM/\lambda }$
             & $138\,\mathrm{km/s}$
             & $0.2\, c$ \\
Time                     & $\sqrt{\lambda^3/(GM)}$
             & $0.58\, \mathrm{days}$
             & $0.51\, \mathrm{days}$ \\
Mass density             & $\mathcal{B}^2\lambda/(\mu_0 GM)$
             & $4.2 \times 10^{-14}\, \mathrm{g/cm^3}$
             & $1.8 \times 10^{-17}\, \mathrm{g/cm^3}$ \\
\hline
\end{tabular}
\end{center}
\caption{\protect \small{The magnetic
permeability is denoted by $\mu_0$. 
Representative values are listed for a young stellar object (YSO) with
$M = 1\, M_{\protect\sun}$, $\mathcal{B} = 10\, \mathrm{G}$, and $\lambda = 10\,
\protect R_{\protect\sun}$. Likewise, an entry is included for a typical active galactic nucleus
(AGN) of $M = 10^8\, M_{\protect\sun}$, $\mathcal{B} = 100\, \mathrm{G}$ and $\lambda =
10\, R_\mathrm{S}$, where the Schwarzschild radius is given by $R_\mathrm{S} =
2GM/c^2$.}}
\label{tab:units}
\end{table*}

Both for numerical reasons and to be able to adapt the calculations
to stellar as well as galactic scales, the physical quantities are
given in units of characteristic quantities of the system.
The form of the equations are not changed by converting to this
system of units. The appropriate units may be constructed from
the mass of the central object, $M$, the magnetic flux
density, $\mathcal{B}$, and the length scale, $\lambda$. The characteristic
length scale $\lambda$ is assumed to correspond to the inner disc radius.
If not otherwise stated, the dimensionless quantities listed in Table
\ref{tab:units} are used in the following sections.

The code uses a sixth order accurate method for partial derivatives and
a fifth order accurate interpolation method with the variables
represented in non-uniform (Eulerian) staggered meshes
\cite{code,Rognvaldsson99}.
In the discrete representation of quantities, the scalar variables ($e$ and
$\rho$) are zone centered and components of the vector variables ($\mathbf{p}
= \rho \mathbf{u}$ and $\mathbf{B}$) are face centered in a unit mesh volume.
The solution for the eight variables, $\rho$, $e$, $p_x$, $p_y$,
$p_z$, $B_x$, $B_y$, $B_z$ is advanced in time by a third order
predictor-corrector procedure \cite{hyman}, modified to accommodate
variable time steps.
The code has previously been verified by a number of standard test
problems and henceforth been used
for experiments involving 3-D turbulence \cite{Nordlund+94a},
investigations of problems related to coronal heating \cite{Nordlund+Galsgaard96spm},
buoyant rise of magnetic flux tubes \cite{Dorch+Nordlund97buoy},
dynamo experiments \cite{Dorch+99a},
stellar convection \cite{Nordlund+Stein99budapest}
and magnetized cooling flows \cite{Rognvaldsson+00a}.

Cartesian coordinates ($x, y, z,$) are used to describe the disc-corona
system in the computational box. To calculate the derivative or the
interpolated value at a given point, the six nearest grid points are
involved. As exactly the same operators are used in the whole grid,
three layers of \emph{ghost zones} must be added to prevent
periodic wrapping of the computational domain. For (obsolete) reasons of
computational efficiency, we use $x$ (first index) as the vertical direction
and place ghost zones in the index range $i \in [1:3] \wedge [m_x-3:m_x]$.
The $y$- and $z$-directions are taken to be periodic.
The ($y,z$) cross section of the computational domain is quadratic,
both in terms of number of grid points ($m_y = m_z$) and physical size ($L_y =
L_z$), and centered so that $y \in [-L_y/2,L_y/2]$ and $z \in [-L_z/2,L_z/2]$.
Only odd numbers of $m_y$ and $m_z$ have been chosen, making $(x,y,z)=(0,0,0)$
at zone center $(i,j,k)=(4,(m_y + 1)/2, (m_z+1)/2)$.

\subsection{Initial conditions}
\label{sec:ic}

To investigate the magnetic driving and collimation of outflows it is
desirable to choose an initial state in which all other dynamical effects
are small. Therefore, to obtain such a state, the corona is
assumed at rest and in hydrostatic equilibrium in the gravitational field of
a point mass \cite{OP97b}. Self-gravity of the gas is neglected.
A current free configuration is chosen so that the Lorentz force vanishes.
In this way, the momentum equation for the corona reduces to
\begin{equation}
\nabla P = - \rho \nabla \Phi.
\end{equation}
By the use of a polytropic equation of state, $P = K \rho^\gamma$,
the reduced
momentum equation may be integrated, giving the density distribution
for the corona,
\begin{equation}
\rho = \left[ (\Phi_0 - \Phi) \frac{\gamma-1}{K \gamma} \right]^{\frac
{1}{\gamma - 1}}, \quad \gamma \neq 1.
\label{eq:density}
\end{equation}
A positive value of the integration constant, $\Phi_0 \geq 0$,
may be introduced to keep finite densities at large distances.
The value of the polytropic exponent is
chosen to be $\gamma = 5/3$, corresponding to an adiabatic mono-atomic ideal
gas.

For numerical reasons, the gravitational potential is smoothed by introducing
a so-called softening parameter, $l_\mathrm{s}$,
\begin{equation}
\Phi = - \frac{1}{\sqrt{x^2 + y^2 + z^2 + l_\mathrm{s}^2}}.
\label{eq:Phi}
\end{equation}
Due to the location of the gravitational potential in the staggered meshes, a
non-zero softening parameter is necessary to make the potential well-defined
at $x = y = z = 0$. Furthermore, a softened potential
makes the potential gradient and the initial density gradient
less steep close to the central object and therefore easier to handle by the
numerical code. The softening parameter was chosen to be $l_\mathrm{s} =
\sqrt{1/2}$, for the experiments actually reported herein.

The disc surface (lower boundary) is assumed to have a density distribution
described by Eq.~\ref{eq:density}, and the corona is in pressure
balance with the underlying disc. In the case of negligible thermal
pressure,
the softened potential gives rise to an equilibrium Keplerian like velocity
structure in the disc given by
\begin{eqnarray}
u_y &=&   \frac{z}{(y^2 + z^2 + l_\mathrm{s}^2)^{3/4}}, \nonumber \\
u_z &=& - \frac{y}{(y^2 + z^2 + l_\mathrm{s}^2)^{3/4}}.
\end{eqnarray}
Numerical experiments
relying on a smoothed or softened potential produce generally weaker jets
\cite{BL95}.
Due to the reduced rotation velocity obtained when softening is introduced
a longer time scale
for the build up of the azimuthal field (and the jet flow) is anticipated.
Devising identical experiments and only varying softening, we find no
significant effect in jet features but only a scaling of dynamical
quantities. The main effect of softening is an increase of the rotational
period which scales the characteristic time and the specific kinetic
energy (cf.\ Fig.~\ref{fig:discvel}).

In the chosen numerical setup, a density jump at the disc-corona interface
must be applied with some caution, as the interpolation of the zone centered
values could artificially bring mass into the corona. A density jump may
however be implemented to have effect on the energy flux as is
demonstrated in Sect.~\ref{sec:bc} below. The pressure balance between
disc and corona together with the assumed disc density structure expressed
by Eq.~\ref{eq:density} implies that a density jump may equivalently be
regarded as a jump in internal energy per unit mass, $T \sim P/\rho$. Hence,
the jump $\xi \equiv (T_\mathrm{corona}/T_\mathrm{disc})_0$ at the disc-corona
interface expresses a jump in (isothermal) sound speed as well;
$c_\mathrm{s,disc}^2 = c_\mathrm{s,corona}^2 / \xi$.
For $c^2_\mathrm{s,disc} \ll (u^2_y + u^2_z)$, the radial pressure gradient
in the disc, $\partial p / \partial r \sim (\rho c_\mathrm{s}^2)_\mathrm{disc}
/ R$, becomes negligible compared to the centrifugal force and the
pressure gradient does not influence the radial balance of the disc. Anyway,
since the disc is not simulated, the disc structure will not be of a numerical
concern.

A constant vertical magnetic field, $B_x = B_0$, is chosen to penetrate
both disc and corona initially.
In a more realistic case, one may expect the field to be inclined with respect
to the disc surface and the field strength is likely to increase towards
the center due to advection by the accreting matter \cite{CaoSpruit}. Because
of the periodic boundaries, such a magnetic field configuration would
not be divergence free in the present setup. Furthermore,
a field configuration facilitating wind ejection is expected to
be generated automatically by the rotation of the disc \cite{OP97b}.
Therefore, only the simple constant vertical magnetic field configuration
has been implemented as initial condition.

The magnetic flux density is specified by the parameter $\beta_\mathrm{i}
\equiv P_\mathrm{gas,i}/P_\mathrm{mag,i}$, at $r_\mathrm{i}=1$. In the numerical
experiment, the density distribution
is determined from Eq.~\ref{eq:density} with
$K = (\gamma -1)/\gamma$. Using $P_\mathrm{gas} = K \rho^\gamma$ and
$P_\mathrm{mag} = B_0^2/2$
it follows that
\begin{equation}
B_0 = \sqrt{ \frac{2K}{\beta_\mathrm{i}}
\left( \Phi_0 + \frac{1}{\sqrt{1+l_\mathrm{s}^2}}
\right)^{\frac{\gamma}{\gamma - 1}}}.
\label{eq:B0}
\end{equation}

\subsection{Boundary conditions}
\label{sec:bc}

The disc boundary is located in the $x=0$ plane and is assumed to be a fixed base
for the jet. As such, the
rotational velocity, mass density and internal energy density is kept constant
at their initially prescribed values. Jet fluxes of mass, energy and momentum
are assumed to have negligible impact on the disc surface layer. For instance,
the disc is assumed to supply mass to the surface at the same rate as the
jet carries mass away. This is implemented as a \emph{steady} condition where
the mass flux $p_x$ is symmetric at the lower boundary.\footnote{In the
1-D case, a symmetric or zero $x$-derivative condition on the mass flux $p_x$
results in a steady state where
$\partial \rho / \partial t = - \partial p_x / \partial x = 0$ at
the boundary.} The energy flux at the disc surface has been implemented as
either a zero x-derivative condition or a cold condition. The cold condition
is applied by assuming a jump in
internal energy per unit mass (temperature) between disc and corona. The
internal energy density in the disc ghost zones are found by antisymmetrizing
the $x$-face centered values of $T \equiv e/\rho$ around $(e_0/\rho_0)/\xi$;
\begin{equation}
T[b-i+1/2] = 2 (e_0/\rho_0)/\xi - T[b+i-1/2],
\end{equation}
with $b=4$ and $i=1,2,3,4$ so that the values in square brackets are the grid
locations. The fixed internal energy density and mass
density at the disc surface is denoted $e_0$ and $\rho_0$ respectively.
The temperature jump is specified
by the ratio $\xi=(T_\mathrm{corona}/T_\mathrm{disc})_0$ evaluated at the
disc-corona interface. In effect, a cold inflow condition is specified
by the internal energy density flux, $h_x \equiv p_x T$.

The inflow velocity is not specified, but allowed to evolve freely according
to the forces acting on the disc surface. However, the $x$-flux of $x$-momentum
(the $xx$-component of the Reynolds stress tensor) is not allowed to change the
outflow velocity on the surface. Instead, thermal and magnetic forces
are taken to be the dominant mechanisms for injecting gas into the corona.
The pressure in the disc (ghost zones) and on the boundary is kept fixed at the
values prescribing the initial pressure balance. Any pressure gradient at the
disc surface is caused solely by deviations from the initial
hydrostatic equilibrium occurring in the corona.

For the evolution of the magnetic field the disc is taken to be perfectly
conducting. In general, the
magnetic field is only specified initially and is
subsequently evolving according to the electric field. Keeping this in
mind when imposing boundary conditions, violation of the divergence free
condition of the magnetic field may be prevented in a natural way.
Particularly, one finds that
the disc velocities determine the electric field and
thereby the evolution of the magnetic field at the lower boundary. The electric
field on the disc surface is specified as
\begin{eqnarray}
E_y & = & - u_z B_0, \nonumber \\ E_z & = & u_y B_0,
\end{eqnarray}
where $B_0$ is the constant
vertical field specified initially by Eq.~\ref{eq:B0}.
The issues of boundary driving is discussed further in Sect.~\ref{sec:bd}.

At the upper boundary, an extrapolation of the electric field would not be
stable and another approach has to be applied which is discussed in
Sect.~\ref{sec:qtb}.
For the mass, momentum and energy density fluxes, a simple extrapolation
into the ghost zones may be applied to allow the densities to evolve on the
boundary.

\subsubsection{Boundary driving}
\label{sec:bd}

At the lower boundary the disc is kept rotating with a fixed angular
velocity. This should in principle be sufficient to determine the
evolution of the $y$- and $z$-components of the magnetic field.
It turns out, though, that the velocity driving has to be specified in a
somewhat non-local manner. This is because of the non-local
nature of the numerical differential operators used in the
code. The numerical scheme involves the
three nearest points on each side along the direction of differentiation.
Therefore, to avoid ripples when driving, the driving has to be passed
smoothly to the active grid. To be specific, the velocity field just inside
the active grid is determined by passing a third order polynomial through
the boundary and the two neighboring zones further inside the grid,
\begin{equation}
u_i[5] = \frac{4}{9} u_i[4] +
         \frac{6}{9} u_i[6] - \frac{1}{9} u_i[7], \quad i=y,z,
\end{equation}
where the lower boundary is located at zone index 4.
The above chosen polynomial representation has been found in other
but similar experiments to give the smoothest driving \cite{code}.
It has not been subjected to tests in the present work.

\begin{figure}
\psfig{figure=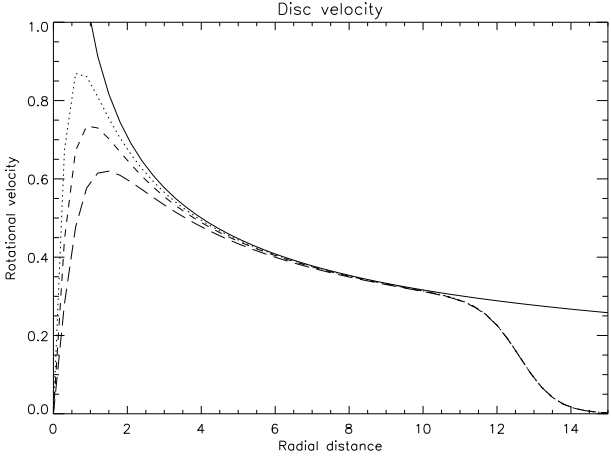,width=8.8cm}
\caption{\protect \small{Plots of the rotational velocity in the disc for
different values of the softening parameter, $l_\mathrm{s}$.
The solid, dotted, dashed and long dashed lines
represent values of $l_\mathrm{s} = 0$ (Keplerian), $1/2$,
$\protect \sqrt{1/2}$ and 1 respectively.
The plotted profile is given by,
$u(r) = u_{\theta}f$ where $u_{\theta}= r/(r^2+l_\mathrm{s}^2)^{3/4}$.
The cutoff function $f$ terminates the rotation at the periodic boundary
($r \protect \gtrsim 10$) but does not affect the inner regions of the disc.}}
\label{fig:discvel}
\end{figure}

Care has to be taken also to avoid ripples caused by the periodicity of the
$y$- and $z$-boundaries and by the shape of the box. To avoid shear at the box
sides and reversed vorticity at the box corners, the disc velocity profile
has to be terminated inside the periodic boundaries. As above, the non-local
nature of the differential operators demands the decrease of disc rotation
to span a few grid points. For the same reason, the velocity cutoff has
to take place at some distance from the boundaries to ensure a region
of practically zero rotation. To satisfy these conditions,
a hyperbolic tangent function is used to cut off the velocity;
\begin{equation}
f_i(r) = \frac{1}{2}\left[ \tanh \left( \frac{L_i}{2}-s_i-r \right) +1 \right],
\quad i=y,z.
\label{eq:velcut}
\end{equation}
Here $L_i$ is the size of the box in the $i$ direction and $s_i$ is the
distance that the midpoint of the hyperbolic cutoff is shifted from
the boundary.  The resulting velocity profile, $u_{i,\mathrm{cut}}=u_i f_i$,
is shown in Fig.~\ref{fig:discvel} for $L_y = L_z = 30$ and $s_y = s_z =
2.5$ in a slice along $y=0$ and $z \ge 0$.

\subsubsection{The quasi-transmitting boundary}
\label{sec:qtb}

A symmetric condition on the electric field does not
allow tangential components of the magnetic field to evolve on the boundary.
In particular, when the initial winding of the vertical magnetic field
reaches the upper boundary as a toroidal Alfv\'{e}n wave, it would be reflected.
To minimize reflection (reversal of the toroidal magnetic field) and
possible disruptive and/or decelerating effects on the flow, 
an upstream sensing of changes in the magnetic field is used to guide the
evolution of the boundary field. This may appear somewhat artificial, but it
corresponds roughly to apply values along an outgoing characteristic
of a linear torsional Alfv\'{e}n wave. Such a wave propagates with the
Alfv\'{e}n speed $u_\mathrm{A} = B_x/\sqrt{\rho}$. As a
fair estimate, changes in the tangential field are taken to propagate with the
(fixed) Alfv\'{e}n speed determined by
\begin{equation}
u_{\mathrm{A},x} = \frac{B_0}{\sqrt{(\Phi_0 - \Phi)^{\frac{1}{\gamma - 1}}}},
\label{eq:uadv}
\end{equation}
evaluated at the upper boundary.

The tangential magnetic field on the boundary is assumed to evolve towards some
upstream value at a distance $\Delta x_\mathrm{up}$, below the boundary.
The travel time is estimated from the Alfv\'{e}n speed; $t_\mathrm{up} =
\Delta x_\mathrm{up} / u_{\mathrm{A},x}$. The condition implemented on the
electric field specifies the evolution of the tangential magnetic field as
\begin{equation}
\frac {\partial B_i}{\partial t} =
\frac {\Delta B_{i,\mathrm{up}}}{t_\mathrm{up}}, \quad
i=y,z.
\end{equation}
The work done on the field may be controlled by the
parameter $\chi$, which specifies the
convergence values for the magnetic field at the upper boundary;
\begin{equation}
\Delta B_{i,\mathrm{up}} = B_i[m_x-4] - \chi B_i[m_x-4-n_x], \quad i = y, z.
\label{eq:Brise}
\end{equation}
Here, $n_x$ specifies the index width of the sensing distance.
Choosing $\chi \lesssim 1$ mimics in a simple way the work done on the magnetic
field by the part of the jet outside the computational box.

It must be emphasized, that the $\chi$ parameter is to be
regarded as a real physical parameter expressing to what extent the exterior
is ``braking'' the field rotation. If, for example, the field is anchored to
massive regions in the ambient medium outside the computational domain, the
braking will cause the field to become inclined with respect to the outer
boundary. Hence, the choice of $\chi$ is not only a question of numerics, but
reflects a real physical degree of freedom.

\subsection{Note on approach}

The model philosophy adopted here is different from previous work,
particularly in the applied boundary conditions. Instead of using large
efforts on designing (artificially) open outer boundary conditions,
periodic boundaries have been applied. The argument being, that it is
inconceivable that real jets would not be perturbed by motions
in the surrounding medium anyway. In the present work, the outer boundaries as
well as numerical noise provide the perturbations triggering the
possible instabilities we wish to investigate. Obviously, this approach
cannot provide precise determinations of e.g.\ growth rates, but it is
used to allow instabilities to develop in a natural way and provide
indications on whether our understanding of jet collimation needs to
be revised or not.


\section{Results}
\label{sec:results}

The numerical experiments all show the same overall evolutionary sequence
which may be summarized as follows:
\begin{enumerate}
\item   Winding and opening of the magnetic field lines.
\item   Relaxation of initial magnetic acceleration and spurious
    magnetic reflections.
\item   Build up of collimated (but unsteady) outflow.
\end{enumerate}

Initially, the winding of the magnetic field propagates through the
computational box as a torsional Alfv\'en wave. The winding propagates with
velocities expected to vary with distance approximately as Eq.~\ref{eq:uadv}.
Accordingly the winding propagates faster at large radii as noted also by
Ouyed and Pudritz \cite*{OP97b}.

The upper boundary condition on the magnetic field allows (partially) for
the Alfv\'en pulse to be transmitted. Some reflections are unavoidable, but
eventually the reflections are transmitted through the boundaries and the system
settles down.
The relaxation of the minor Alfv\'en pulse reflections concludes the two initial
phases of the numerical experiments.
At which evolutionary stage a (significant) collimated outflow is initiated
in a real accreting system can only be speculated upon. In the present type of jet
experiment, the initial phases do not reveal much relevant physics and
the initial conditions as well as boundary conditions are probably not well suited
for such investigations. However, the setup eventually provides a collimated
outflow with the expected qualities that we want to investigate.

A set of parameters have been introduced to truncate the physical
problem and adopt it into a numerical setup. To probe the effect of
these parameters, such as softening, box size and velocity cutoff,
a range of experiments were carried out. The truncation of the
problem and the effect of the related parameters have been presented
above. In what follows, we will concentrate on a small number of experiments
which have been conducted particularly to investigate stability issues. These
experiments are listed in Table \ref{tab:experiments}.

\newcounter{subrun}
\newcounter{run}
\renewcommand{\therun}{\emph{\roman{run}}}
\begin{table}[h!bt]
\begin{center}
\begin{tabular}{c||r|r|r|r|l}
\hline
run &
$ m_x \times m_y \times m_z$ &
$ L_x \times L_y \times L_z$ &
$\beta_\mathrm{i}$ &
$\xi$ &
$t_\mathrm{max}$ \\
\hline
\hline
\refstepcounter{run} \label{run:largejet1} \therun &
$98\times71\times71$ & $40\times30\times30$ &
 5 & 
 1 & 500 \\
\refstepcounter{run} \label{run:largejet2} \therun &
$98\times71\times71$ & $40\times30\times30$ &
 10 & 
 1 & 500 \\
\refstepcounter{run} \label{run:largejet4} \therun &
$98\times71\times71$ & $40\times30\times30$ &
 5 & 
 100 & 500 \\
\refstepcounter{run} \label{run:hugejet1} \therun &
$128\times85\times85$ & $60\times40\times40$ &
 10 & 
 1 & 400 \\
\refstepcounter{run} \label{run:hugejet2} \therun &
$200\times101\times101$ & $200\times25\times25$ &
 10 & 
 1 & 260 \\
\hline
\end{tabular}
\end{center}
\caption{\protect \small{List of experiments for the stability study. All
experiments use the softening parameter, $l_\mathrm{s} = \protect \sqrt{1/2}$,
and a polytropic index, $\gamma = 5/3$.}}
\label{tab:experiments}
\end{table}

\subsection{Jet dynamics}
\label{sec:dynamics}

By examining the rates of change of energies,
the action of various forces and issues of stationarity may be analyzed.
The
rate of change of
total energy in a volume of gas is
controlled by the rate of energy transport in and out of the volume,
i.e.\ the net flux, $\Delta F=F(x_\mathrm{lower})-F(x_\mathrm{upper})$, the rate of
energy conversion through work, $W$, and dissipation, $Q$. Consequently, the
rate of change of magnetic, $\mathcal{M}$, kinetic, $\mathcal{K}$, thermal,
$\mathcal{T}$, and gravitational energy, $\mathcal{G}$, in a volume may be
expressed as follows,
\begin{equation}
\frac{\partial \mathcal{M}}{\partial t} =
 \Delta F_\mathrm{mag} + W_\mathrm{Lorentz} - Q_\mathrm{Joule},
\end{equation}
\begin{equation}
\frac{\partial \mathcal{K}}{\partial t} =
 \Delta F_\mathrm{kin} + \Delta F_\mathrm{visc} +
   W_\mathrm{grav} - W_\mathrm{gas} - W_\mathrm{Lorentz}
 - Q_\mathrm{visc},
\end{equation}
\begin{equation}
\frac{\partial \mathcal{T}}{\partial t} =
 \Delta F_\mathrm{enth} + W_\mathrm{gas} + Q_\mathrm{visc} + Q_\mathrm{Joule},
\end{equation}
\begin{equation}
\frac{\partial \mathcal{G}}{\partial t} =
 \Delta F_\mathrm{grav} - W_\mathrm{grav}.
\end{equation}
The rate of total work done by gravity in the box, $W_\mathrm{grav}$,
the rate of work done against gas pressure gradients, $W_\mathrm{gas}$,
and the rate of work done against the Lorentz force, $W_\mathrm{Lorentz}$,
are given by
\begin{equation}
W_\mathrm{grav} = - \int_V \mathbf{u} \cdot \rho \nabla \Phi \, dV,
\end{equation}
\begin{equation}
W_\mathrm{gas} = \int_V \mathbf{u} \cdot \nabla P \, dV,
\end{equation}
\begin{equation}
W_\mathrm{Lorentz} = - \int_V \mathbf{u} \cdot \left(
\mathbf{J \times B} \right) \, dV.
\end{equation}
Here $dV \equiv dx\,dy\,dz$ is an element of the volume, $V = [x_\mathrm{lower},
x_\mathrm{upper}] \times [-L_y/2,L_y/2] \times [-L_z/2,L_z/2]$ . The energy
fluxes through a cross section, $x$, are given by
\begin{equation}
F_\mathrm{mag}(x) =
 \int_S (E_yB_z - E_zB_y) \, dS_x, \label{eq:flux1}
\end{equation}
\begin{equation}
F_\mathrm{kin}(x) =
 \int_S u_x \left( \frac{1}{2} \rho u^2 \right) \, dS_x,
\end{equation}
\begin{equation}
F_\mathrm{enth}(x) =
 \int_S \rho u_x \gamma e \, dS_x,
\end{equation}
\begin{equation}
F_\mathrm{grav}(x) =
 \int_S \rho u_x \Phi \, dS_x, \label{eq:flux2}
\end{equation}
where $dS_x \equiv dy \, dz$ represents the element of a surface normal to
the $x$-direction and the surface, $S = [-L_y/2,L_y/2] \times [-L_z/2,L_z/2]$,
is an entire cross section of the box.

\begin{figure*}
\vspace{0cm}
\hbox{\hspace{0cm}\psfig{figure=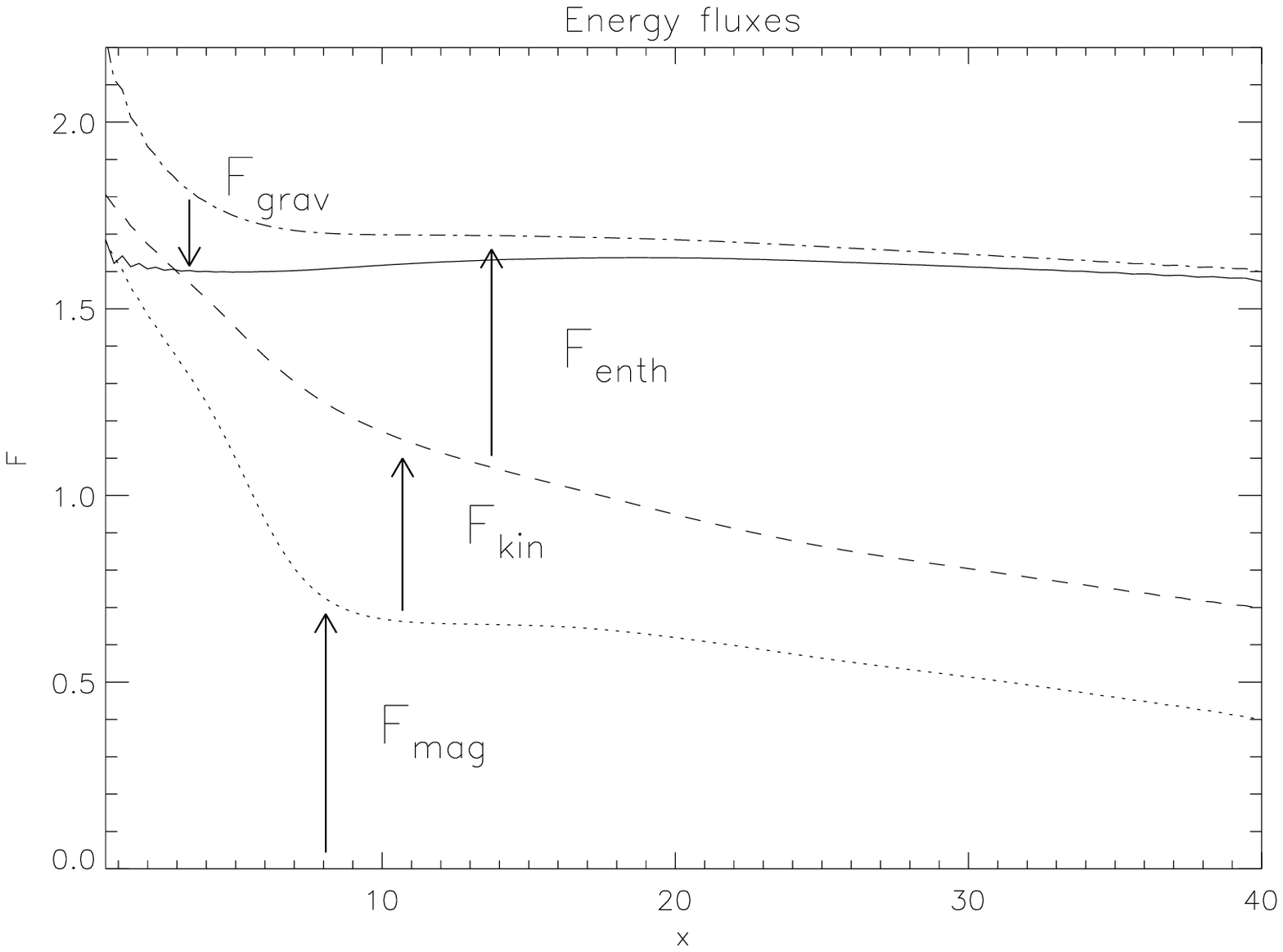,width=8.8cm}\hspace{0cm}
\psfig{figure=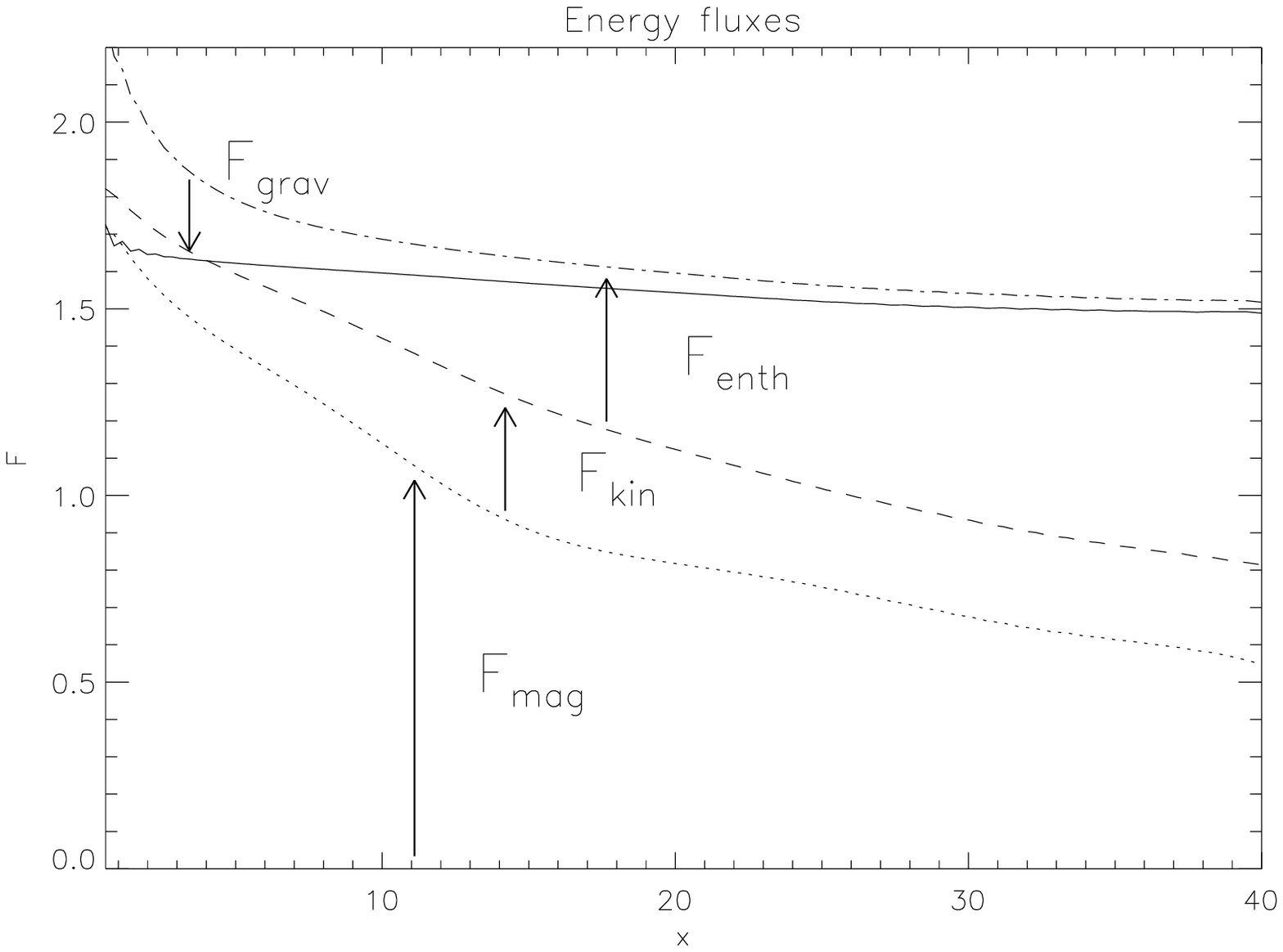,width=8.8cm}}
\vspace{0cm}
\caption{\protect \small{Time averaged total energy fluxes as functions of
height for run \ref{run:largejet1} (left panel) and run \ref{run:largejet2}
(right panel). The dotted line is magnetic flux,
dashed line is the sum of kinetic and magnetic fluxes, the dot-dashed line is
the sum of enthalpy, kinetic and magnetic fluxes. The solid line is the
total energy flux, which is obtained by adding the (negative) flux
of gravitational energy to the sum of the other fluxes (dot-dashed line).}}
\label{fig:totalflux}
\end{figure*}

To demonstrate the quasi-stationary state of the flow and the general energy
conversion mechanisms, the time averaged energy fluxes (Eqs.~\ref{eq:flux1}--\ref{eq:flux2}) are plotted in Fig.~\ref{fig:totalflux} as functions of
height above the disc. The fluxes are obtained as differences between the
curves as indicated by the arrows in the plots.

For both experiments it is seen that more magnetic energy is transported into
the volume than is carried out. The magnetic energy does not increase accordingly,
since the difference between magnetic energy flux in and out of the volume
is balanced on average by Lorentz work and Joule dissipation.
The conversion of magnetic energy flux into
kinetic energy flux is dominant close to the disc, whereas magnetic
energy flux is primarily converted into thermal energy flux in the upper two thirds
of the box for both the experiments. Just above the disc surface, a
(hydrostatic) balance is seen to be maintained between gravitational energy
flux and the sum of enthalpy and kinetic energy flux. This corresponds to a
situation where gravity is balanced by gas pressure and $x$-advection of
momentum. The two
experiments differ significantly in the amount
of magnetic energy flux converted into thermal energy flux as a consequence
of the difference in the magnetic energy reserve at hand.

\begin{figure*}[h!tb]
\mbox{\psfig{figure=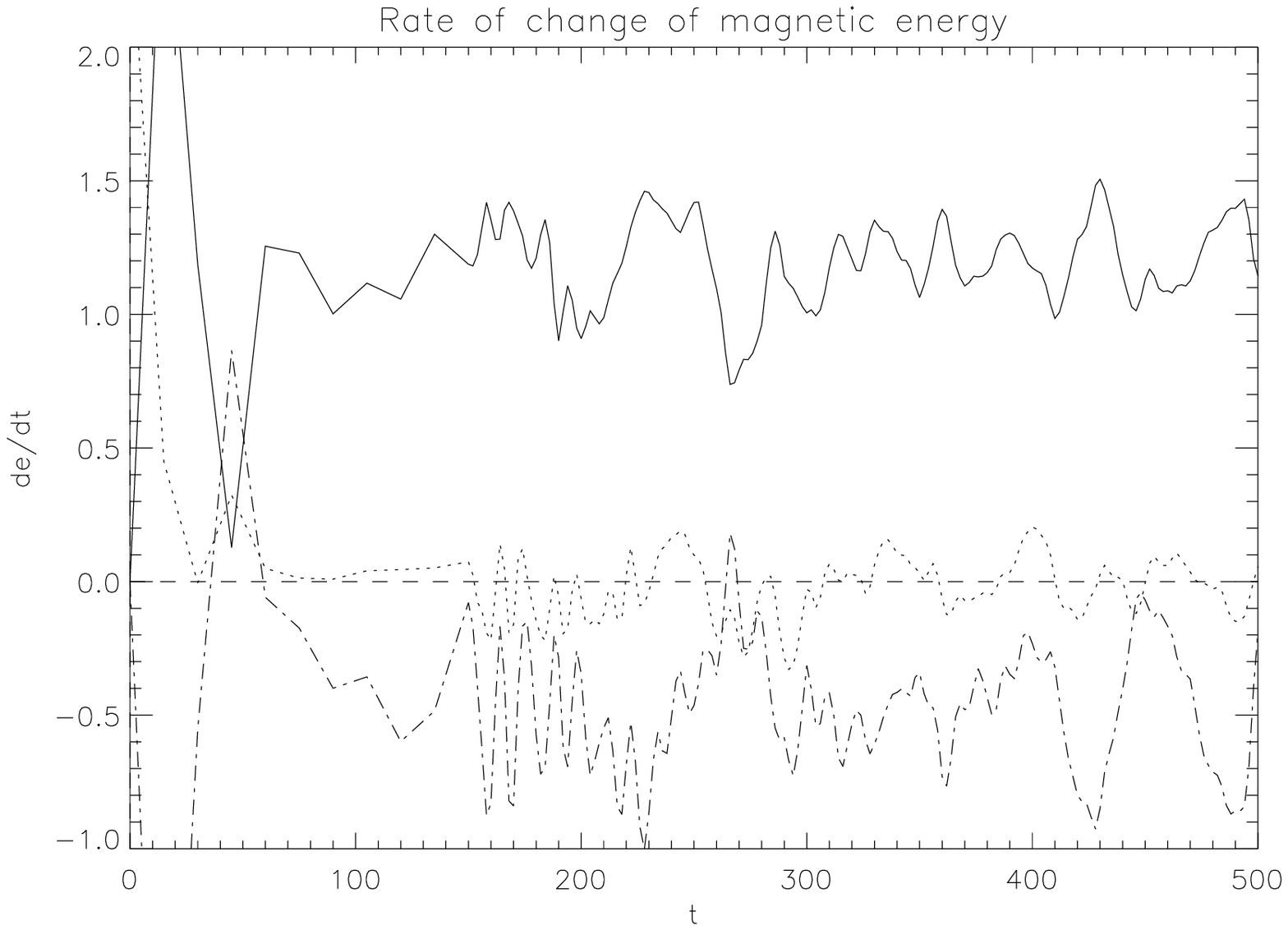,width=8.8cm}\hspace{0cm}
\psfig{figure=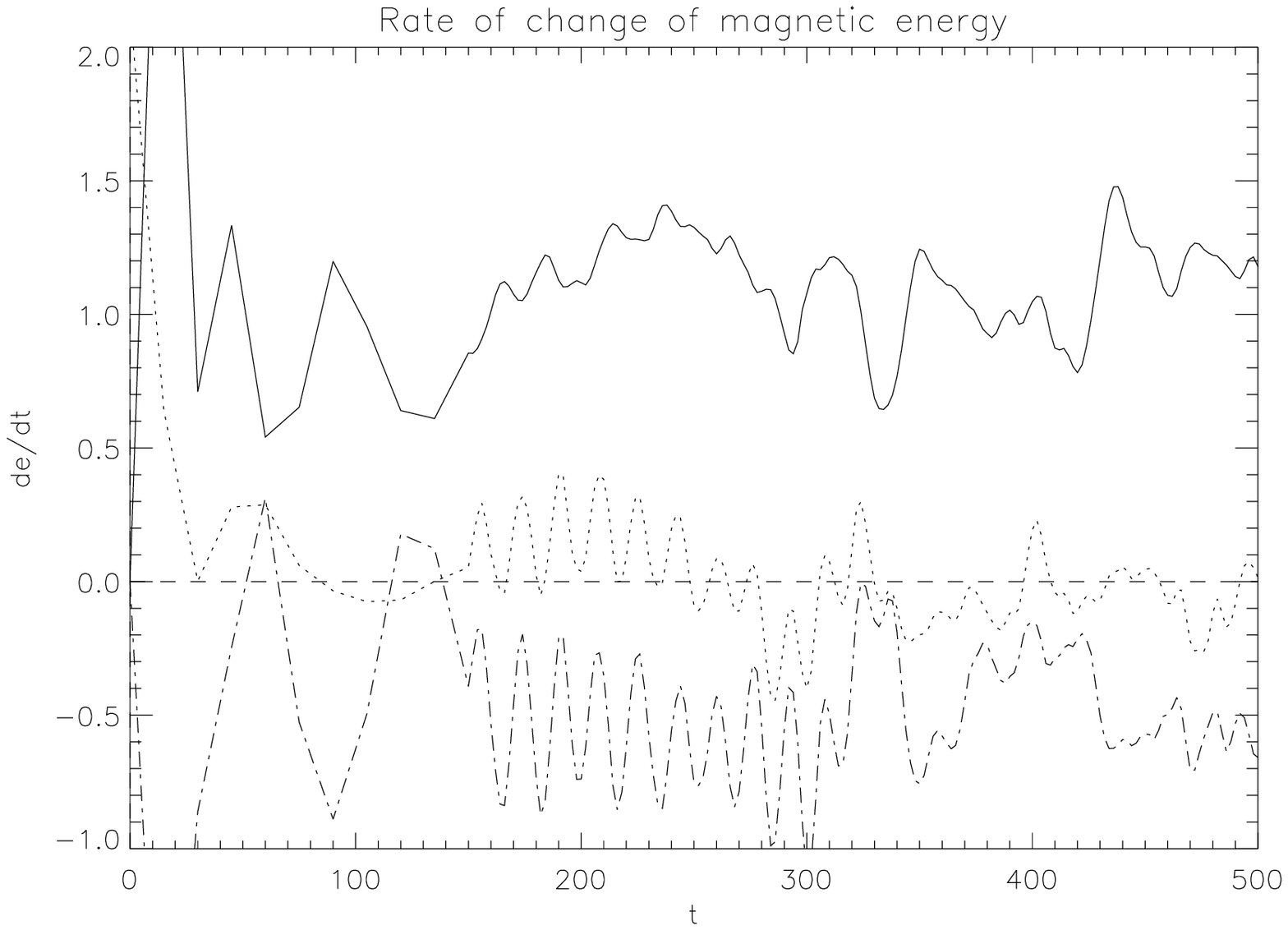,width=8.8cm}}\\
\vspace{0.2cm}
\mbox{\psfig{figure=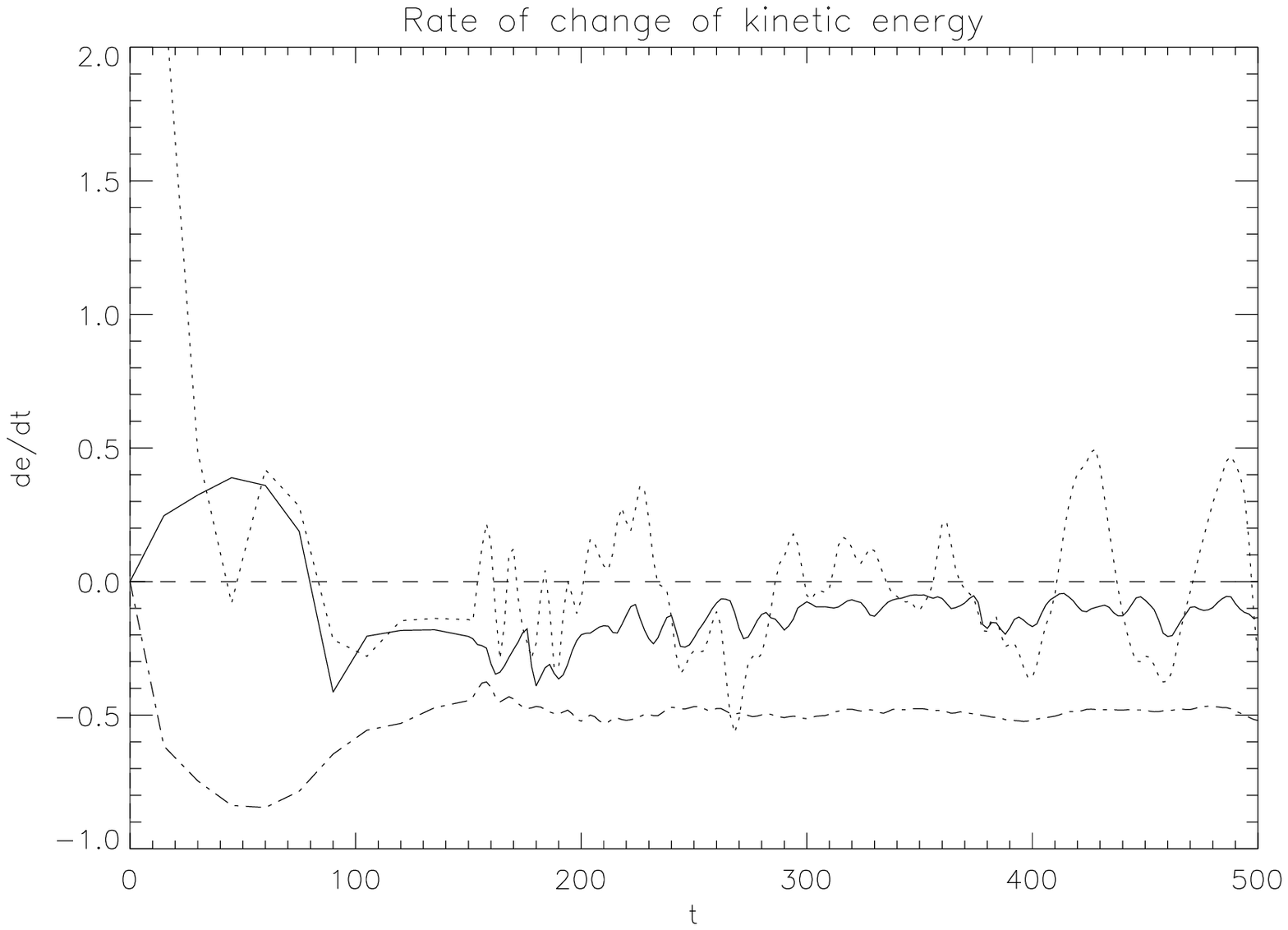,width=8.8cm}\hspace{0cm}
\psfig{figure=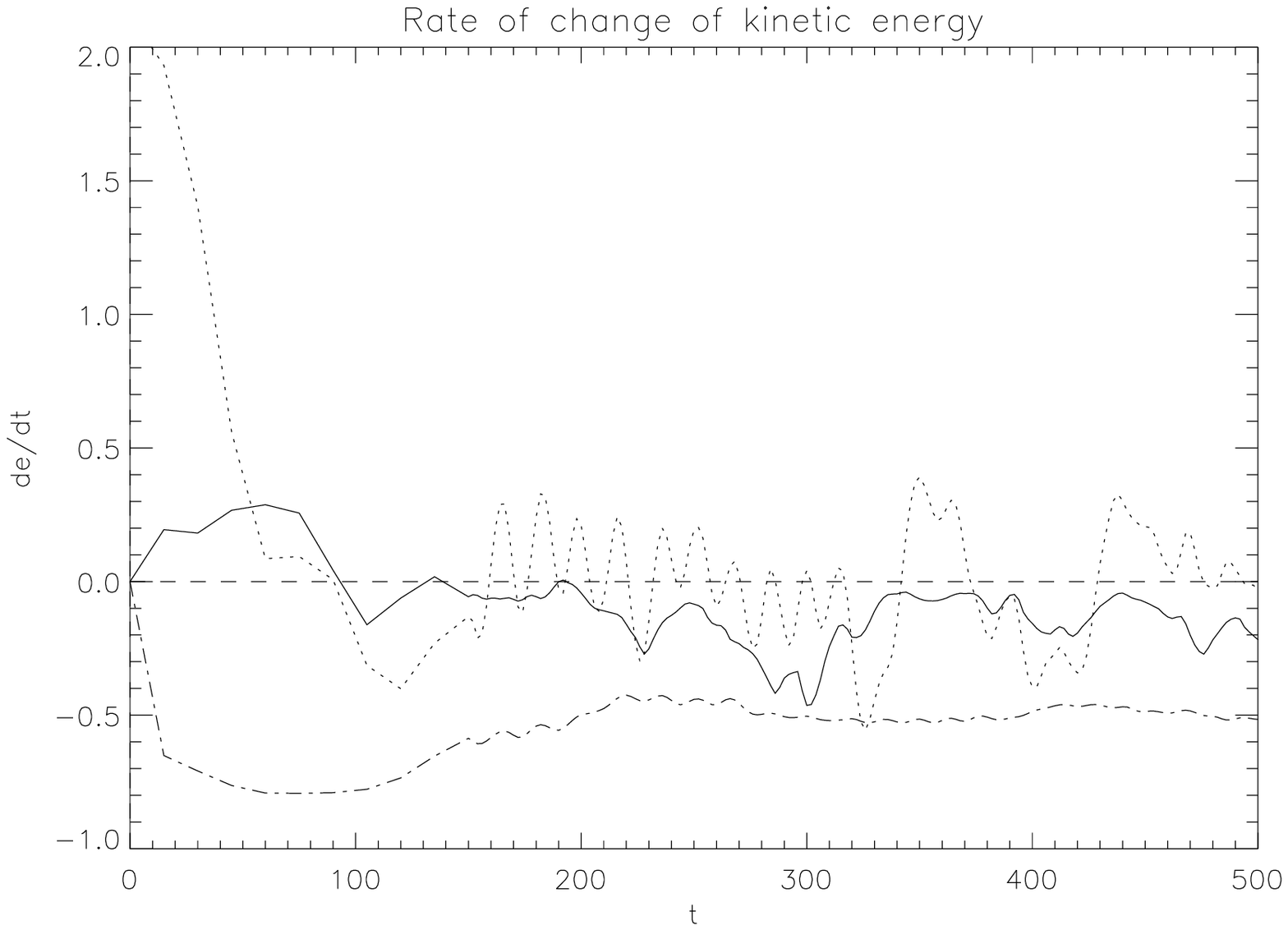,width=8.8cm}}\\
\vspace{0.2cm}
\mbox{\psfig{figure=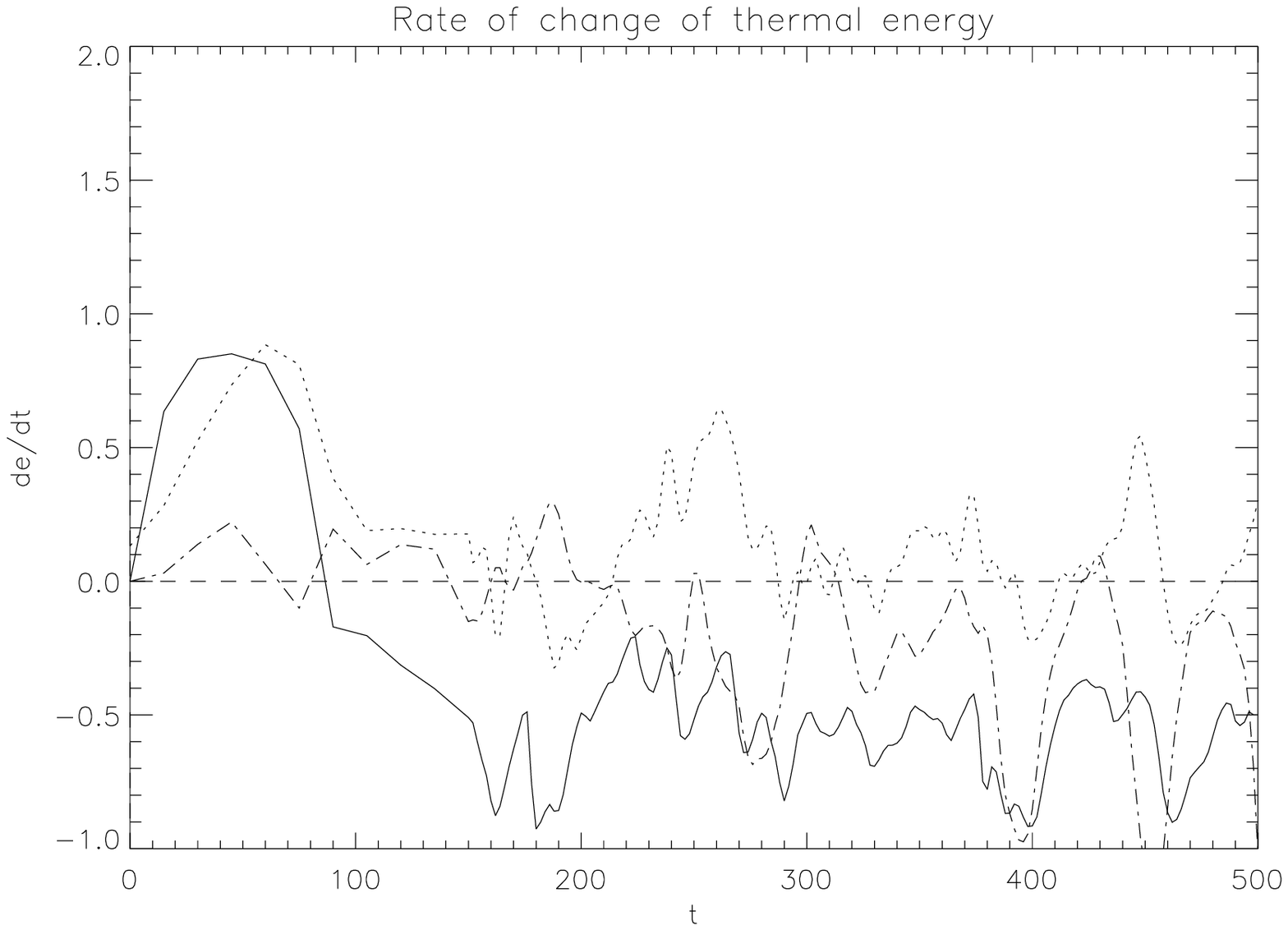,width=8.8cm}\hspace{0cm}
\psfig{figure=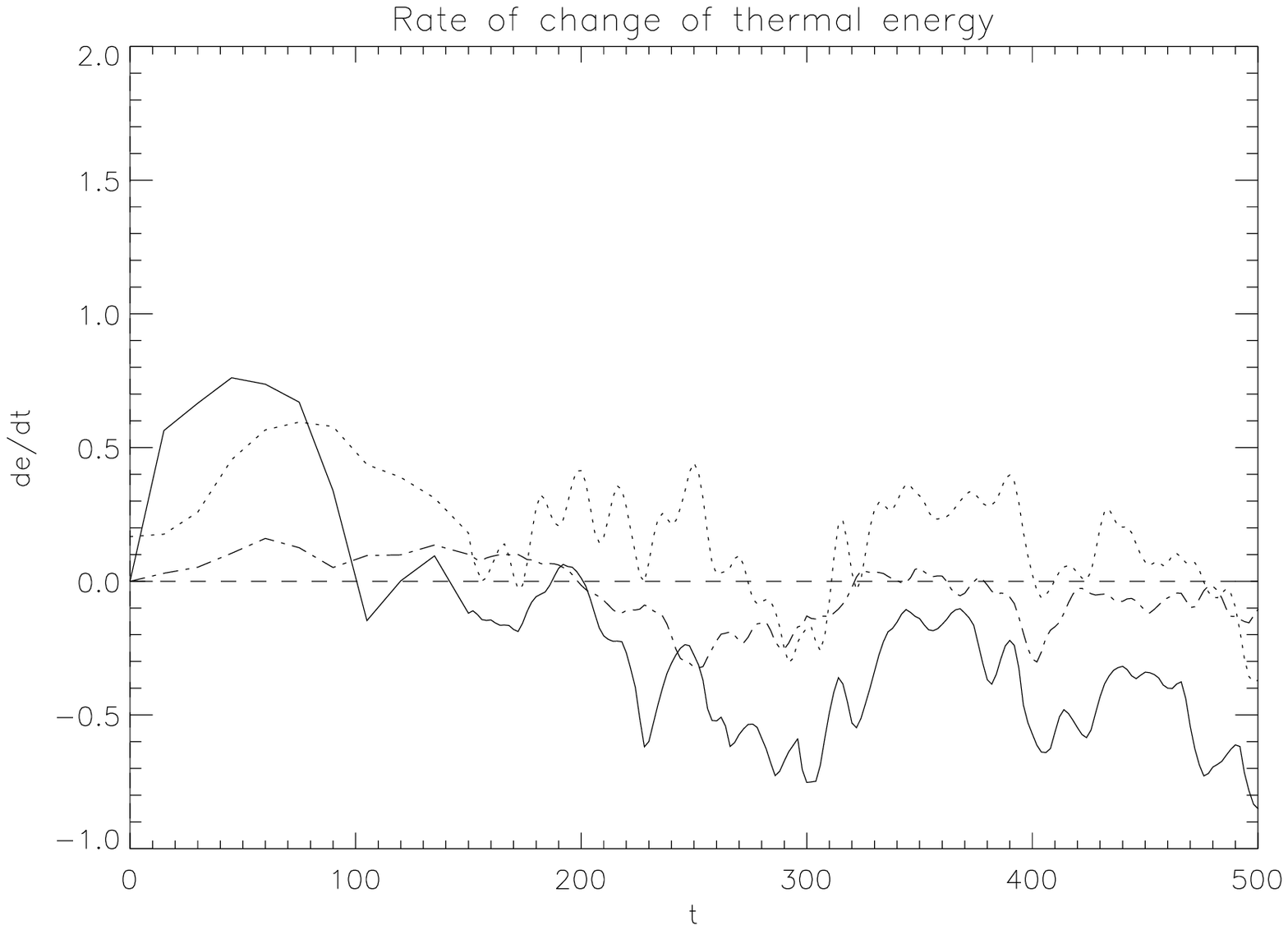,width=8.8cm}}\\
\caption{\protect \small{The rate of change of magnetic, kinetic and thermal
energy in the box between $x=1.1$ and $x=38.8$ are graphed for run
\ref{run:largejet1} (left panels) and run \ref{run:largejet2} (right panels).
The top panels show rate of change of magnetic energy (dotted line),
the net Poynting flux into the volume (solid line) and the rate of
work done by the Lorentz force (dot-dashed line).
The middle panels show rate of change of kinetic energy (dotted
line), net kinetic flux into the volume (solid line) and the rate of
work against gravity done by carrying matter through the volume
(dot-dashed line). The lower panels show rate of change of thermal energy in the
volume (dotted line), net enthalpy flux (solid line) and the rate of work
done by gas pressure (dot-dashed line). The dashed line marks zero rate of
change of energy. Please note that until $t=150$ only 10 snap shots were
saved (at time intervals $\Delta t = 15$).}}
\label{fig:largejet1-dedt}
\end{figure*}

Details of the energy conversion processes as a function of time are shown
in Fig.~\ref{fig:largejet1-dedt}. After a sharp increase of
kinetic energy, caused by the initial Alfv\'{e}n pulse, the rate of change of
kinetic energy oscillates as does the rate of work done by the Lorentz force.
Both the efflux of kinetic energy and the work done by gravity are fairly
constant, but large variations in the work done by gas pressure are seen
for run \ref{run:largejet1}. The characteristics of the events at the
``pressure work peaks'' ($t \approx 270, 400, 450$) for
run \ref{run:largejet1} are all indications of major, dynamically important
magnetic reconnection events.  Such events relax the
wound up magnetic field configuration, whereby Lorentz work is reduced
as there is no longer the same amount of azimuthal magnetic field for
driving (and pinching) the flow. Consequently, the rate of change of kinetic
energy drops. Before this happens, the thermal
energy has been building up for some time (positive rate of change of
thermal energy) and much of this energy is now released by pressure work. This
promptly makes the rate of change of kinetic energy positive. As
the magnetic pinching is significantly reduced,
one possible effect of the pressure work ``explosions'' is to generate
filaments and disruptions of the jet into the ambient medium. The
magnetic pressure of the surrounding vertical magnetic field
eventually halts this (irregular) radial expansion (cf.\
Fig.~\ref{fig:collforces}). The
events at $t \approx 400$ and $t\approx 450$ are followed by an increase in the
rate of Lorentz work which supports this picture.
As a consequence of the work done by gas pressure, and the increased enthalpy
flux out of the volume following the reconnection events, the rate of change of
thermal energy becomes negative.
The slight decrease in net Poynting flux is in line with a
constant Poynting flux at the lower boundary but a decreased Poynting flux
out of the volume. The sudden deficit in Poynting flux through the upper
boundary is a consequence of azimuthal field relaxation in the box, causing
the tangential field components to decrease at exit.

Comparing the two experiments run \ref{run:largejet1} and run
\ref{run:largejet2} the weak field experiment (run \ref{run:largejet2})
appears much less violent with respect to magnetic reconnection events.
The most prominent events to be identified occur relatively late, at
$t \approx 320$ and $t \approx 400$. Apparently these events do not cause
significant changes in the jet dynamics, as no abrupt peaks in the pressure
work are detected.
Instead, another prominent feature of the jet dynamics may be
identified, namely the oscillatory pattern in the rate of Lorentz work and
kinetic energy flux. Such harmonic oscillations (in the radial flow) between
an inner ram pressure barrier (centrifugal barrier) and an outer magnetic
pressure barrier have been predicted analytically \cite{SautyTsinganos} and
shown in axisymmetric experiments to steepen into fast magneto-sonic shocks
\cite{OP97b}. The oscillations are most clearly present until approximately
the time where the first prominent reconnection event takes place and
complicates the flow pattern.

\subsubsection{Forces and flow features}
\label{sec:features}

\begin{figure}[h!tb]
\psfig{figure=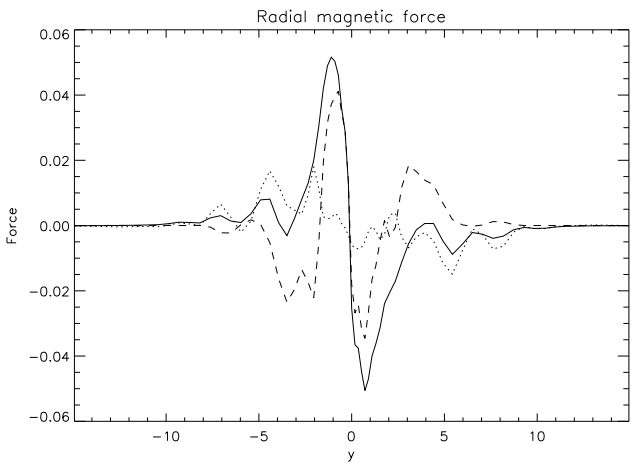,width=8.8cm}
\caption{\protect \small{The collimating magnetic forces at $x \approx 4$
for run \ref{run:largejet2} at $t=200$. The dashed line is the radial (here the
$y$-direction) pressure
gradient of the toroidal magnetic field (here, $B_z$) and the
dotted line is the radial pressure gradient of the vertical magnetic field.}}
\label{fig:collforces}
\end{figure}

A look at the forces reveals that
close to the central object the gas is launched from the disc surface
by a thermal pressure gradient between disc and corona. Just
above the disc the acceleration process may be regarded as either
magnetic and centrifugal. Close to the rotation axis the magnetic point of
view may conveniently be adopted as the magnetic field is here highly wound
up and a vertical gradient of the field is the main acceleration
mechanism. At larger radial distances, the magnetic field is
not too wound up for the gas to be flung more radially outwards. Here the
acceleration mechanism bares the same characteristics as in the
centrifugally driven wind scenario.
The central jet region appears very well collimated from the start, as the
magnetic acceleration process for this region is inherently vertical and
causes little radial flow. The fraction of the jet flung more radially
outwards is eventually collimated by the magnetic force.
Fig.~\ref{fig:collforces}
shows the collimating magnetic force decomposed into toroidal and radial
magnetic pressure.
The ambient vertical magnetic field has generally a small collimating effect,
but when magnetic reconnection events relaxes the wound up field structure
the vertical field takes over and prevents locally and for a period of
time the flow from expanding into the ambient medium.

The mean vertical velocity increases approximately linearly with height
and the terminal velocities obtained in the jet at the upper
boundary are of the order of the Kepler velocity at the inner disc radius.
However, there is some dependence of the mean terminal velocities on the
strength of the initial field as expected from the predictions by Michel
(Michel, 1969; see also Kudoh et al., 1997).\nocite{Michel,Kudoh}
The time averaged terminal velocities are consistent
with a power law dependency, $u_\infty \propto B_0^{2/3}$,
in the relatively early quiescent stages of the jet evolution.
At later stages, disruptive events become dynamically important and the
assumptions for the predicted terminal velocity dependency, e.g.\ that the
gas dynamics is governed purely by magnetic effects, do not hold. The more
violent disruptive events in the strong field experiment redirect a
relatively larger fraction of the vertical flow into transverse gas motions.
Eventually, the mean vertical velocity of the strong field experiment becomes
less than the mean vertical velocity of the weak field experiment.

\begin{figure}
\psfig{figure=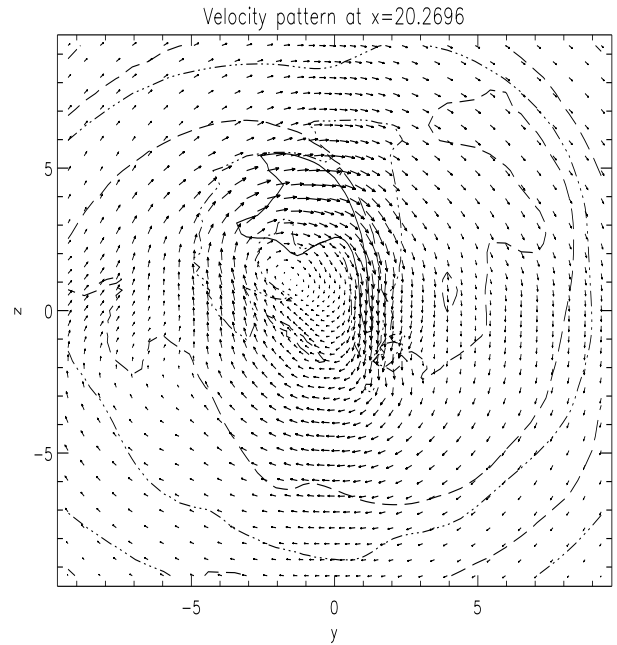,width=8.8cm,height=8.8cm}
\caption{\protect \small{Complexity of the flow illustrated in a cross
section at $x \approx 20$ for \ref{run:largejet1} at $t=270$. The dashed line
contours mark zero vertical velocity, dash-dot-dot-dot line contours mark
the sonic points and the solid line contour marks the Alfv\'{e}n surface.}}
\label{fig:flow}
\end{figure}

To illustrate the complexity of the resulting flow motions,
Fig.~\ref{fig:flow} shows velocity vectors in a cross section of the flow.
Contours indicating the Alfv\'{e}n surface, super-sonic regions and
regions of back-flow are drawn. The super-sonic and super-Alfv\'{e}nic
jet beam is surrounded by a supersonic shell at larger radii. Furthermore,
a prominent back-flow region is noted just outside the jet beam in the
upper right quadrant of the plot.

\begin{figure}[h]
\psfig{figure=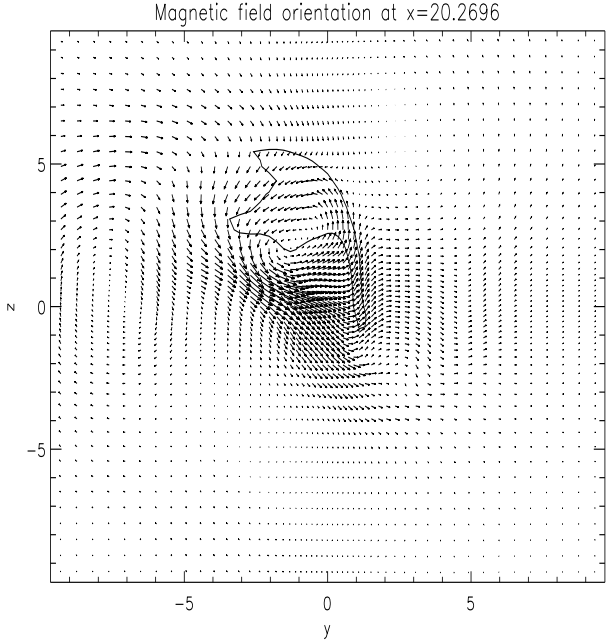,width=8.8cm,height=8.8cm}
\caption{\protect \small{The magnetic field orientation for run
\ref{run:largejet1} at $t=270$ in a cross section at $x \approx 20$. The
contour line marks the Alfv\'{e}n surface.}}
\label{fig:unwind}
\end{figure}

\subsection{Jet stability}
\label{sec:stability}

\begin{figure*}[t!hb]
\begin{center}
\mbox{\psfig{file=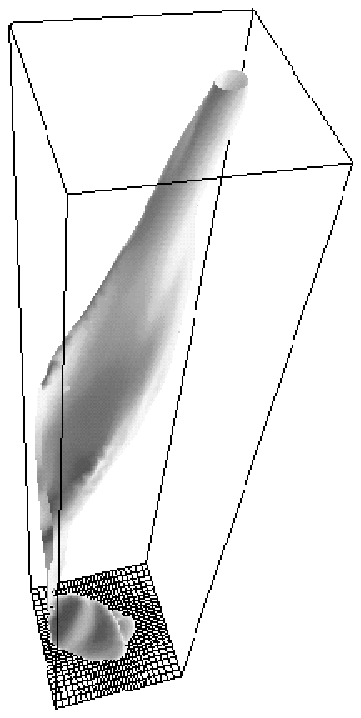,width=5cm}\hspace{0cm}
\psfig{file=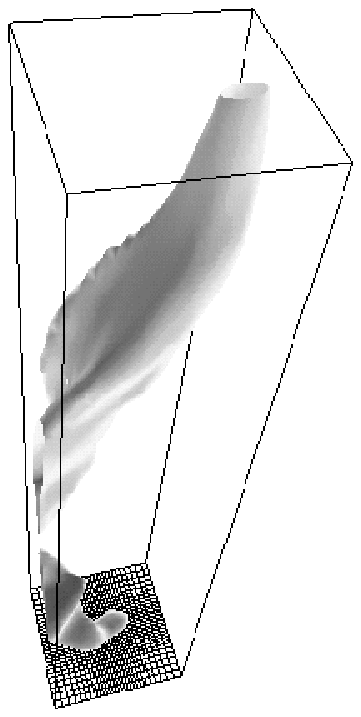,width=5cm}\hspace{0cm}
\psfig{file=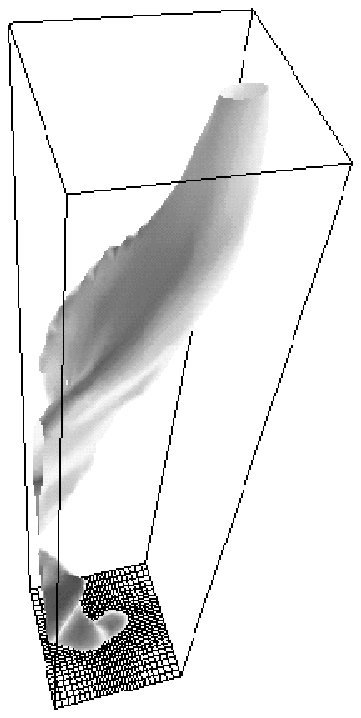,width=5cm}}
\vspace{0cm}
\mbox{\psfig{file=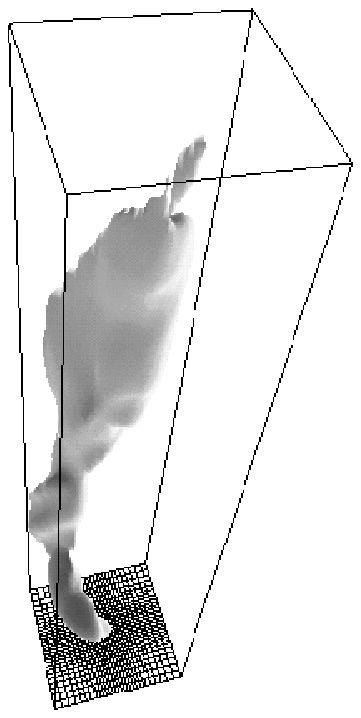,width=5cm}\hspace{0cm}
\psfig{file=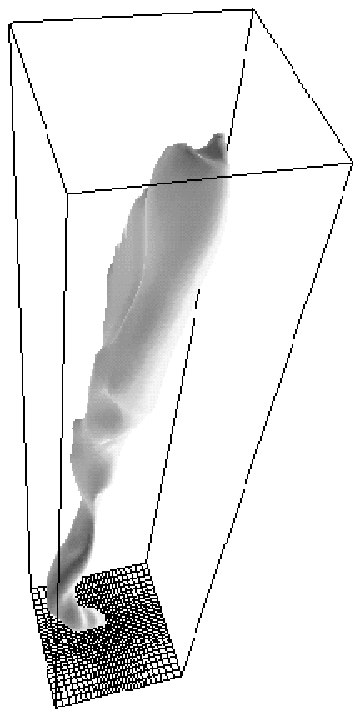,width=5cm}\hspace{0cm}
\psfig{file=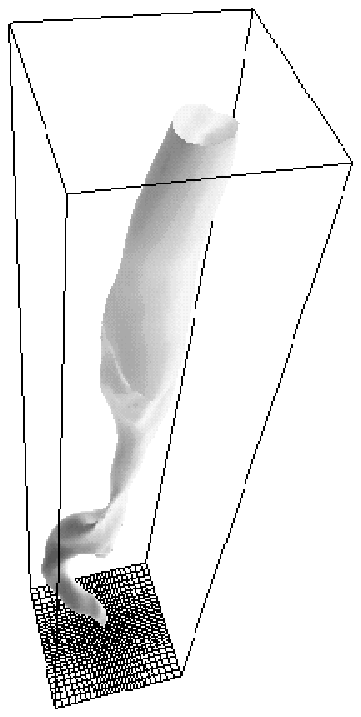,width=5cm}}
\end{center}
\caption{\protect \small{Isosurfaces of magnetic flux density for
run \ref{run:largejet1}. The snap shots are taken at $t=270, 274, 278, 282,
286, 290$ starting at the upper left panel and ending at the lower right.
The middle section of the flux tube is located off center and partly outside
the test volume at first, but
the non-axisymmetric distortion relaxes and the tube swings (clockwise)
into a more axisymmetric configuration with a tightly wound spiral
structure close to the disc surface. For clarity, only a narrow region of the
computational domain centered around the $x$-axis is shown.}}
\label{fig:kink1}
\end{figure*}

Though potentially disruptive events may be identified in the dynamics, these
events are found only to generate filaments and cause non-destructive
distortions of the jet beam. To monitor the onset and evolution of
instabilities further, the magnetic field topology is investigated.
Phenomenological investigations of
the evolution of the field topology are carried out by visualizing isosurfaces
of magnetic flux density.
Fig.~\ref{fig:kink1} shows the evolution of an isosurface of the magnetic
flux density. The magnetic field is seen to develop a
complex and highly non-axisymmetric structure, which relaxes periodically
allowing the field to return to a less complex, more nearly cylindrical
topology. In each such build-up/relaxation cycle several modes may be
identified qualitatively such as the sausage instability and the elliptical
($|m|=2$) modes.

\begin{figure}[h!]
\psfig{figure=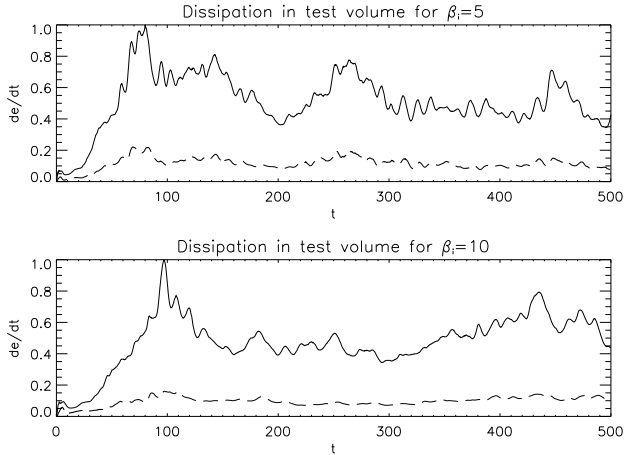,width=8.8cm}
\caption{\protect \small{Total dissipation in a test volume for run
\ref{run:largejet1} (upper panel) and run \ref{run:largejet2} (lower panel).
The dissipation is integrated in a central volume of the
box between $x=[0.7, 38.1]$ and $y,z=[-4.7, 4.7]$ and is
normalized by the maximum value of the magnetic dissipation in each
experiment. The solid lines are magnetic dissipation and the dashed lines
are viscous dissipation.}}
\label{fig:jet1+2-dissip}
\end{figure}

The total viscous and magnetic dissipation as a function of time are shown in
Fig.~\ref{fig:jet1+2-dissip} for a test volume. In the initial phase the wound
up field structure is building up and the magnetic dissipation is seen to
increase sharply until $t \approx 100$.
From the magnetic dissipation, several minor and a couple of major
magnetic reconnection events are identified to occur in this
region for run \ref{run:largejet1} (upper panel).
The evolution of the magnetic field shown in Fig.~\ref{fig:kink1}
is seen to be connected to a significant increase in the magnetic dissipation.
A slight increase is seen also in the viscous dissipation, indicating that
a small fraction of the energy released by field re-organization is transferred
into kinetic energy and subsequently dissipated.

The dissipation of the magnetic field does not match the Poynting flux into
the test volume in general. The surplus of magnetic energy is partly used for
driving the jet flow, as demonstrated in Sect.~\ref{sec:dynamics}, and
partly to build up the winding between relaxation events. In addition, there
is a continuum of ``background'' magnetic dissipation between the relaxation
events which may have important consequences for the jet stability. The
significance of the magnetic diffusion varies in correspondence to e.g.\ the
characteristic flow velocity as. Specifically, one
expects in the $\beta_\mathrm{i} = 10$ case (run \ref{run:largejet2}) the
characteristic (vertical) jet velocity to be less than in the
$\beta_\mathrm{i} = 5$ case (run \ref{run:largejet1}) as noted in
Sect.~\ref{sec:features}.  Accordingly, the magnetic Reynolds number,
$Re_\mathrm{M} = UL/\eta$, will be smaller in the weak field
experiment and magnetic diffusion relatively more important.
The magnetic diffusion is better capable of counter balancing the continual
field winding and the build-up phase of the azimuthal field is prolonged.
For the same reason the relaxation events themselves appear less violent, as
seen by comparing the panels of Fig.~\ref{fig:jet1+2-dissip}, and will
cause less destructive, more localized distortions of the flow.

The damping of the spiraling jet motion at heights where the jet is in general
super-Alfv\'{e}nic is related to what appears as field ``unwinding'' just
outside of the super-Alfv\'{e}nic central beam. Fig.~\ref{fig:unwind}
shows the orientation of the magnetic field in a cross section at $x \approx 20$
and illustrates the unwinding. The jet beam
swings clockwise in the plot, and the change in field orientation is seen
particularly on the front of the Alfv\'{e}n surface in the direction of
the (clockwise) spiral motion.

\begin{figure*}[t!hb]
\begin{center}
\mbox{\psfig{file=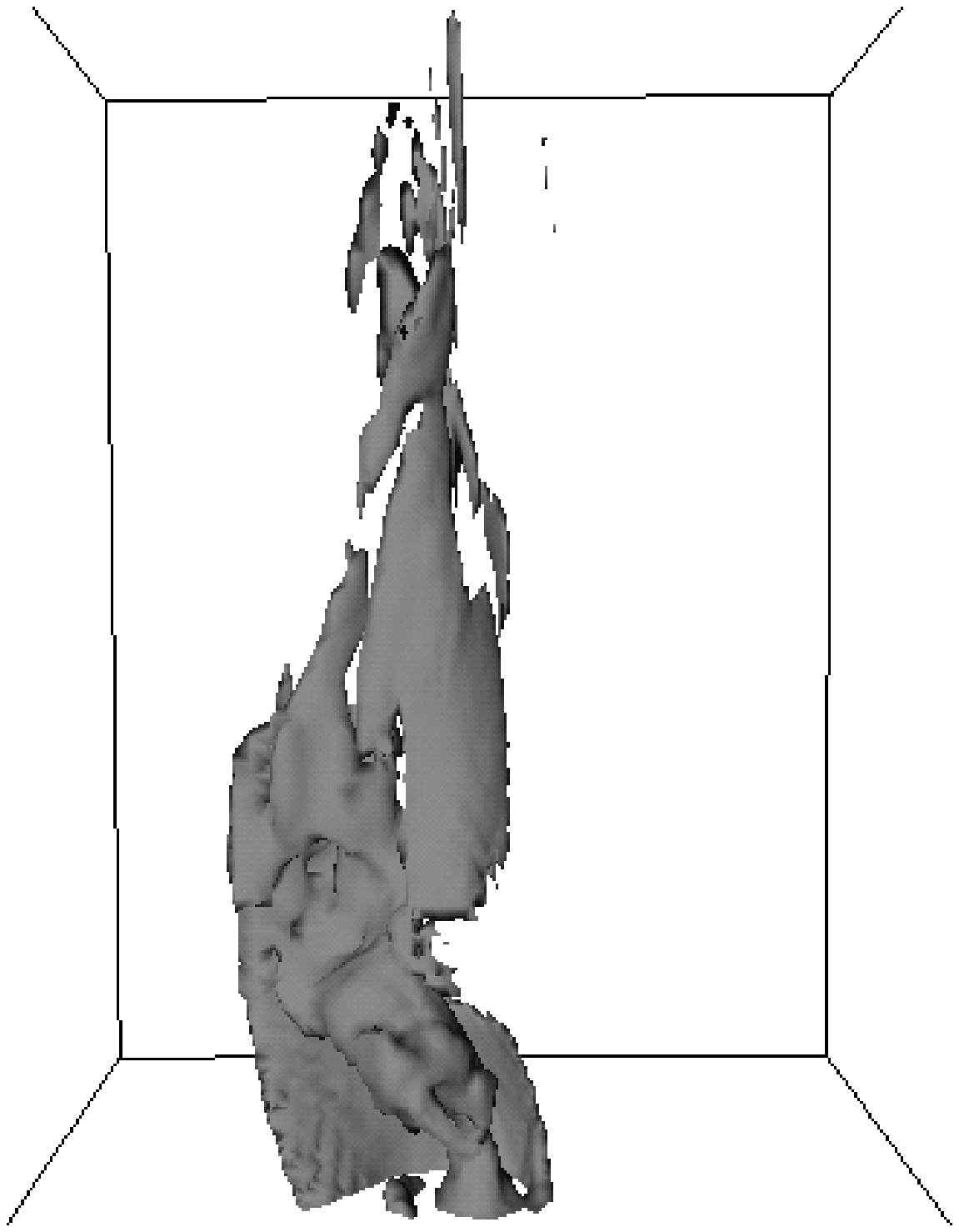,width=8.5cm,height=8.5cm}\hspace{0cm}
\psfig{file=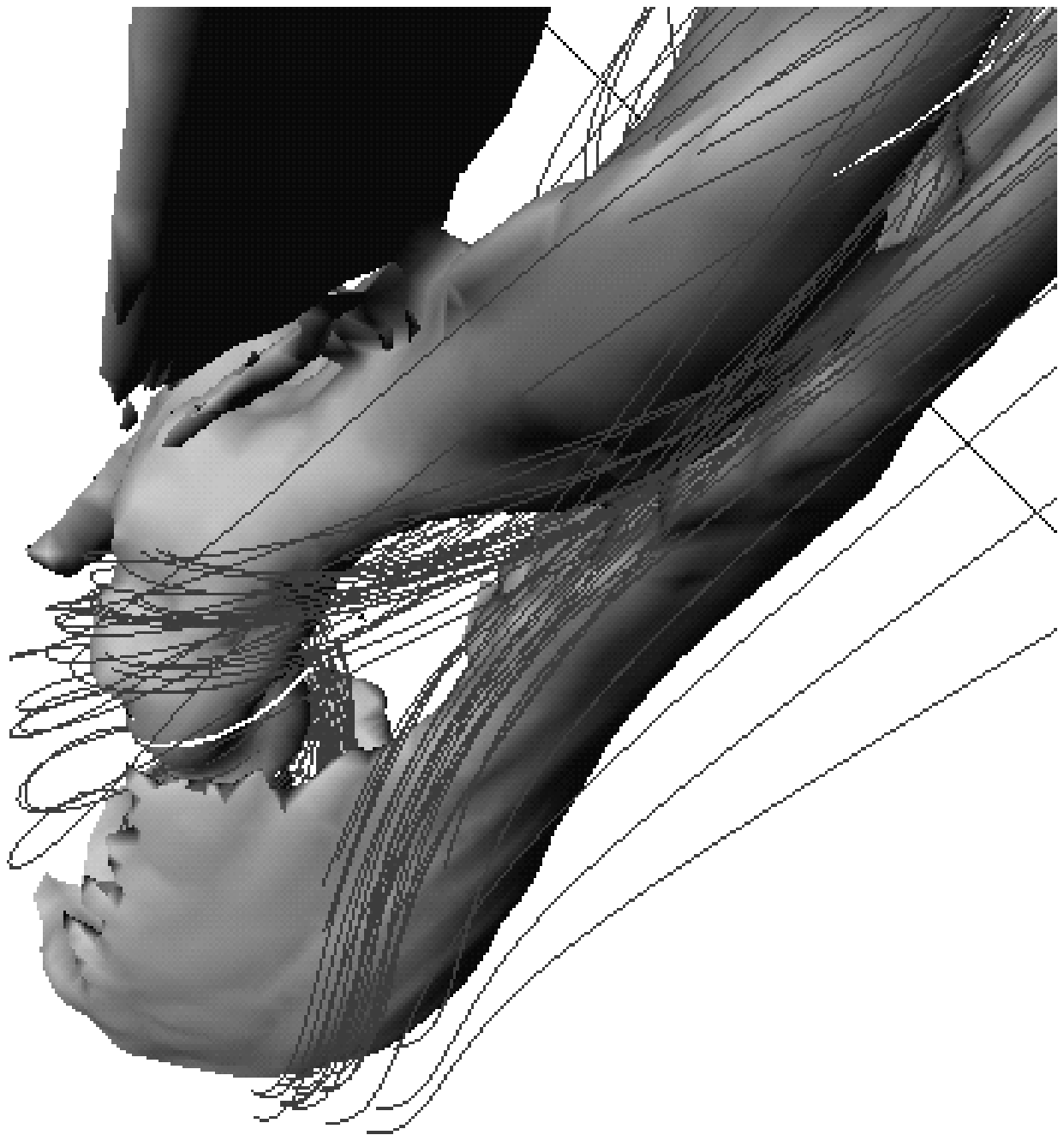,width=8.5cm,height=8.5cm}}
\end{center}
\caption{ \small{The left hand side panel shows isosurface of magnetic dissipation.
The dissipation is seen to occur predominantly in two regions;
in the central highly wound field region close to the rotation axis and in
a ``cocoon'' at the jet flanks.
The right hand side panel shows zoom-in on magnetic dissipation isosurfaces
and inner wound up
magnetic field field lines pressed towards the outer more vertical field. Both
snap shots are from run \ref{run:largejet1} at $t=300$.}}
\label{fig:zoomins}
\end{figure*}

The unwinding of the field comes about as
the spiral motion of the jet beam forces the wound up field into the ambient
almost vertical field. The wound up field is oriented in a direction
practically perpendicular to the ambient vertical field and reconnection
occurs when these oppositely directed field lines are forced into
a small region. This is seen as sheets of Joule dissipation marking the regions
of magnetic diffusion at the jet flanks in the left panel of
Fig.~\ref{fig:zoomins}. The right panel of
Fig.~\ref{fig:zoomins} shows such a sheet in more detail and the orientation
of the field lines reconnecting. These magnetic reconnections, occurring in a
``cocoon'' surrounding the central jet beam, are the reason for
the observed back-flow in Fig.~\ref{fig:flow} and the field unwinding
in Fig.~\ref{fig:unwind}. More specifically, the upper right
region of back-flow in Fig.~\ref{fig:flow} corresponds to where
the reconnection events are expected to catapult gas backwards in this
scenario. The gas
at the ``nose'' of the Alfv\'{e}n surface (at $y \approx 2$, $z \approx -1$)
is accelerated oppositely, i.e.\ forward in the rotation direction and upwards.
Also seen in left panel of Fig.~\ref{fig:zoomins} is a central region of
significant magnetic dissipation. This is the region where reconnection occurs
as a consequence of field interlocking resembling the scenario observed in
experiments concerning coronal heating \cite{xc}. The right panel of
Fig.~\ref{fig:zoomins} confirms this interpretation.

\section{Discussion}
\label{sec:discussion}

Obviously, parameter space has not been probed in all detail in the work
presented here. Issues which
clearly need more attention in future work include the magnetic
configuration, thermal conditions and mass loading. Probing other magnetic
configurations would involve a significant change of the numerical code, whereas
the inclusion of detailed cooling expressions would be straightforward
\cite{Rognvaldsson99,Rognvaldsson+00a}.
Magnetic field configurations which fan away
from the jet axis in a more dipole like fashion may provide less stabilization
high above the disc and far from the rotation axis. However, close to the
central parts of the disc the field lines will, in a
potential configuration as suggested by Cao and Spruit \cite*{CaoSpruit},
bend towards the jet axis as a function of height and not appear much
different from the vertical configuration used in this work. The stabilization
and collimation provided by the poloidal field may be reduced if the magnetic
field is produced in a dynamo active disc. In such a scenario, Brandenburg et
al.\ \cite*{Brandenburg+2000} have found that significant amounts of toroidal
magnetic field are
transported into the corona from the outer parts of the disc. This could
seriously influence the stability.

Theoretically, the stability of twisted magnetic flux tubes is expected
to depend on the diameter to length ratio (aspect ratio), such that relatively
long tubes are most unstable. This is strictly
valid only in the ideal magnetohydrodynamics case. In practice, the rate of magnetic
dissipation tends to increase for more narrow tubes, which reduces the importance
of the aspect ratio \cite{kink}. In the work presented here, no systematic
change was found in the appearance of the jet when the aspect ratio of the experiment was
changed (e.g.\ comparing runs \ref{run:largejet2}, \ref{run:hugejet1} and \ref{run:hugejet2}).

As a consequence of magnetic dissipation occurring primarily in two regions
(cf.\ Sect.~\ref{sec:stability}) the jet flow consists of a hot central beam,
with temperature of the order of the virial temperature at the inner disc
radius, and a hot cocoon. 
Hot jet
flanks has been observed in YSO's and are proposed to be generated by
shocks associated with time variability and entrainment in the flow (e.g.\
\object{HH47,} Hartigan et al., 1993). The results
presented in Sect.~\ref{sec:stability}
suggest that magnetic dissipation may be a major additional mechanism for
heating in this region.

%
Preliminary results (run \ref{run:largejet4}) indicate that relatively cold disc
outflow results in less distorted, more well-defined flows and relatively more
dense jets. A similar result has been reported by Hardee et al.\ \cite*{HCR}
who found dense jets to maintain a high-speed spine which prevents disruption
of the internal jet structure though large helical and elliptical distortions
were present. The experiment, as it is, is already
suitable for addressing further which consequences the temperature of the
disc outflow may have on the jet dynamics.
However, further investigations with special emphasis
on the transsonic region close to the disc surface are desirable to investigate
the issues of mass loading and disc-jet interactions in general.


\section{Summary and conclusions}
\label{sec:conclusion}

A high order numerical scheme has successfully been adopted and a suitable
mesh refinement specified. Initial conditions resembling
previous axisymmetric numerical experiments have been chosen to ease
comparison. The polytropic equation of state
is only used initially, to prescribe the initial density distribution of the
corona.
The most important features that differ from previous jet experiments are:
\begin{itemize}
\item The model is three dimensional rather than axisymmetric. Due to
the periodic boundaries, a cutoff of the disc at large radii is applied.
\item The thermal energy equation is solved, with self-consistently computed
heating by viscous and Joule dissipation.
\item ``Free'' mass outflow from the disc, i.e.\ the mass flux is allowed
      to adjust self-consistently to the forces near the disc surface.
\item Parameterized Poynting flux through the upper boundary, representing
external work.
\item The experiments have been evolved far beyond the initial transient, and
display a quasistationary behavior, as evidenced for example by the nearly
constant total energy flux.
\end{itemize}

The jet dynamics has been investigated by analyzing the mechanisms
of energy conversion. The rotational energy of the disc is carried by the
magnetic field into the corona and is first predominantly converted into kinetic
energy. In the upper two thirds of the computational domain the magnetic energy
is predominantly converted into thermal energy.
General features predicted by steady state theory and axisymmetric numerical
experiments, such as knot generation and terminal velocity dependency on
the magnetic field strength have been confirmed in the relatively early and
quiescent stages of the experiments. At later stages
the flow becomes quite unsteady as instabilities set in, but no serious
disruption of the flow occurs. The jet stability is found to be influenced by
the magnetic dissipation --- this has not previously been investigated in the
context of jet flows. The main findings concerning magnetic dissipation are
the following:
\begin{itemize}
\item The heating by magnetic dissipation is significant, and leads to jet
      temperatures of the order of the virial temperature of the innermost
      Kepler orbit.
\item Magnetic reconnection occurs primarily in two regions: in a central
      region of the jet
      due to field interlocking and in a jet cocoon due to the spiraling motion
      of the jet beam forcing the wound up field into the ambient vertical
      field.
\item Magnetic dissipation helps to prevent critical kinking, and the jet
      is able to sustain a quasistationary flow, with only limited excursions.
\item Reconnection events are seen to result in mass ejection into the
      ambient medium and cause filament structures in the jet beam.
\item Reconnection events are seen to significantly influence the dynamics
      at the jet flanks where deceleration and even back-flow can be found quite
      close to the central super-Alfv\'{e}nic beam.
\item Heating in the region just above the disc is likely
      to have a significant effect on the mass loading.
\end{itemize}


\begin{acknowledgements}
{\AA}N acknowledges partial support by the Danish Research Foundation
through its establishment of the Theoretical Astrophysics Center, Copenhagen.
FT acknowledges the support received from the Theoretical Astrophysics Center,
in granting access to supercomputer facilities both locally and at
UNI-C, {\AA}rhus.

\end{acknowledgements}


\bibliography{aamnem99,aa}
\bibliographystyle{aabib99}

\end{document}